# Spectrally resolved far-fields of terahertz quantum cascade lasers

**Martin Brandstetter**[1,2,*], **Sebastian Schönhuber**[1,2,*], **Michael Krall**[1,2], **Martin A. Kainz**[1,2], **Hermann Detz**[3], **Tobias Zederbauer**[2,4], **Aaron M. Andrews**[2,4], **Gottfried Strasser**[2,4], and **Karl Unterrainer**[1,2]

[1]*Photonics Institute, TU Wien, 1040 Vienna, Austria*
[2]*Center for Micro- and Nanostructures, TU Wien, 1040 Vienna, Austria*
[3]*Austrian Academy of Sciences, 1010 Vienna, Austria*
[4]*Institute of Solid State Electronics, TU Wien, 1040 Vienna, Austria*
*\*Corresponding authors: martin.brandstetter@tuwien.ac.at, sebastian.schoenhuber@tuwien.ac.at*

**Abstract:** We demonstrate a convenient and fast method to measure the spectrally resolved far-fields of multimode terahertz quantum cascade lasers by combining a microbolometer focal plane array with an FTIR spectrometer. Far-fields of fundamental TM0 and higher lateral order TM1 modes of multimode Fabry-Pérot type lasers have been distinguished, which very well fit to the results obtained by a 3D finite-element simulation. Furthermore, multimode random laser cavities have been investigated, analyzing the contribution of each single laser mode to the total far-field. The presented method is thus an important tool to gain in-depth knowledge of the emission properties of multimode laser cavities at terahertz frequencies, which become increasingly important for future sensing applications.

## 1. Introduction

Quantum cascade lasers (QCLs) are compact on-chip laser sources that cover a wide frequency range, including the infrared and the terahertz (THz) spectral regions [1]. Due to the long emission wavelength and the typically small aperture size, engineering the far-field of such devices has been a very active research topic in recent years [2] aiming to provide highly collimated emission for potential applications ranging from imaging to spectroscopy [3]. Several cavity concepts have been demonstrated to achieve this goal, such as photonic crystals [4], DFB lasers [5–7], or non-periodic structures [8,9]. Analyzing the resulting far-field is particularly important for the development and further improvement of different cavity concepts. So far, only spectrally integrated far-field measurements have been performed, either using a single pixel detector [10], which is mounted on a translation stage to scan the emitted radiation at a certain distance, or a focal plane array, enabling real-time acquisition [11]. These techniques are suitable to investigate devices providing single frequency laser emission. However, if multiple modes occur the contributions of each individual mode to the total far-field cannot be distinguished. Since broadband emitting devices are highly desirable for future sensing applications [12,13], a technique enabling spectrally resolved far-field measurements would be beneficial. One particular example are the recently demonstrated surface emitting random laser cavities [14], where each lasing mode shows a different spatial field profile in the cavity which leads to different individual far-field pattern. Decomposing the total far-field into its spectral contributions is therefore very important to further optimize such cavities. One solution is to make use of very narrowband optical filters [15], which, however, suffer from absorption losses and, moreover, require unreasonable long measurement times since a full far-field needs to be recorded for each frequency step. To overcome these issues, here we combine interferometric techniques with a two-dimensional focal plane array to spatially and spectrally resolve the far-field of multimode THz lasers. Similar to a typical FTIR spectrometer, the setup consists of a Michelson interferometer with one scanning mirror and a microbolometer camera as detector. This approach is very sensitive due to the frequency selective nature of the Fourier transformation. Thus, even the far-field of weak lasing modes can be identified unambiguously. Furthermore, due to the real-time acquisition capabilities of the focal plane array, this technique is very fast and requires only a single interferometer scan for a full spectrally resolved far-field measurement. Similar methods provided very convenient solutions to spectrally and spatially analyze the absorption characteristics of various materials at infrared frequencies [16], which even became commercially available recently. Our demonstration shows that this method can further be extended to the THz regime providing a convenient method to analyze the spectrally resolved far-field of multimode lasers.

## 2. Measurement Setup

The setup consists of a commercial FTIR spectrometer (Bruker Vertex 80), where the single pixel detector is replaced by a microbolometer focal plane array (FLIR Tau 2, 324 x 256 pixel), schematically shown in Fig. 1(a). The laser is mounted in a liquid helium flow cryostat and operated at a heat sink temperature of 5 K in pulsed mode with a duty cycle of 4%. To couple the emitted radiation from the laser into the FTIR spectrometer a parabolic mirror with a focal length of 5 cm is used, which allows to measure the far-field within an angle of about 54°. For each detector pixel the interferogram is recorded and a Fourier transformation is performed to extract the spectral as well as the spatial information. The microbolometer camera samples the interferogram resulting from the continuously moving mirror and the laser with a frame rate of 30 Hz. The FTIR spectrometer is operated in free-running mode, where the mirror is moved at its minimum velocity, resulting in interference oscillations of about 12 Hz for a laser emission frequency of 4 THz. Thus, the Nyquist-Shannon sampling theorem, which requires that the maximum measured frequency is smaller than half of the sampling frequency, $f_{max} < f_{sample}/2$, is fulfilled and the measurement can be performed without any synchronization between the camera and the moving mirror in the interferometer. For the highest resolution of 0.075 cm$^{-1}$, one measurement takes approximately 5 min, which is much faster compared to single pixel scanning techniques. For a first test of the system, we compared the spectrum of a laser recorded using the internal single pixel pyroelectric DTGS (deuterated triglycine sulfate) detector (Fig. 1(b), bottom) of the FTIR spectrometer with the spectrum obtained using the microbolometer camera, taking into account the sum of all detector pixels for each frame (Fig. 1(b) top). Both spectra show the same lasing modes. Despite the higher noise level of the camera, the signal-to-noise ratio is sufficient to clearly detect the lasing modes. Better results could be obtained by using a camera which is optimized for the investigated frequency range [17]. Moreover, it should be pointed out that the spectral resolution (in our case 0.075 cm$^{-1}$) only depends on the travelling distance of the moving interferometer mirror and is therefore the same for both measurements.

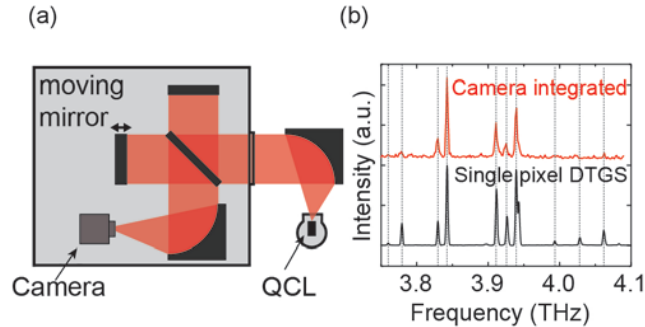

Fig. 1. (a) Scheme of the measurement setup. The QCL is mounted in a liquid helium flow cryostat, and the light is guided via a parabolic mirror into an FTIR spectrometer. The interferogram is recorded either with the internal single pixel DTGS detector or a microbolometer focal plane array (camera). (b) Comparison of spectra obtained by the microbolometer camera (top) and the single pixel DTGS detector (bottom). The same laser modes are obtained with both measurements. Since the resolution of 0.075 cm$^{-1}$ is only determined by the travelling distance of the moving interferometer mirror, it is the same for both measurements.

## 3. Results

### 3.1 Far-Field of a Fabry-Pérot Cavity

A typical resonator for semiconductor lasers is a ridge-type cavity, supporting Fabry-Pérot modes with a frequency spacing of $\Delta f = \frac{c}{2nL}$, with the speed of light in vacuum $c$, the group index inside the cavity $n$ and the resonator length $L$. If the width of such a resonator is wide enough, higher order modes with more than one lateral maximum are supported. For THz QCLs these higher order lateral modes are usually undesired since they reduce the device performance due to a lower confinement within the resonator [18] and lead to a deteriorated far-field. Thus, several techniques have been developed to suppress these modes in order to increase the laser performance and to improve its spectral characteristics. Here, we investigate wide ridge-type cavities without lateral mode suppression in order to force the laser to support fundamental as well as higher order lateral modes. Since the far-field of such mode types is well-known and can be calculated using finite-element methods, this constitutes an ideal system to demonstrate the functionality of the presented spectrally resolved far-field measurement method.

The investigated laser cavity consists of a double-metal waveguide, where the active region is located between two metal layers, confining the radiation in vertical direction. The laser facets are determined by a dry-etching process [19], schematically shown in Fig. 2(e). Simulations of typical Fabry-Pérot modes and the corresponding far-fields of a THz QCL with double-metal waveguide are depicted in Fig. 2(a)-(d). The 3-dimensional simulation geometry consists of the laser active region, modelled by a lossless dielectric (refractive index n = 4.4, 10 µm height, 120 µm width), confined between two metal waveguide layers, represented by perfect electric conducting layers. A commercial finite element solver was used to perform the calculations. Fig. 2(a) and (b) show the distribution of the z-component of the electric field $E_z$ inside the cavity for modes with one (TM0) and two lateral peaks (TM1), respectively. The corresponding far-fields, obtained using the Stratton-Chu method are illustrated in Fig. 2(c) and (d). Since the TM0 mode shows a peak in the center of the far-field ($\varphi = 0$), while the TM1 mode shows a minimum due to destructive interference, they can be clearly distinguished. The black dashed rectangles indicate the fraction of the far-field observable with the used measurement setup, which is limited by the parabolic mirror that couples the emitted radiation into the FTIR spectrometer. Since TM2 and higher order lateral modes have not been observed experimentally, they are not considered in the calculations.

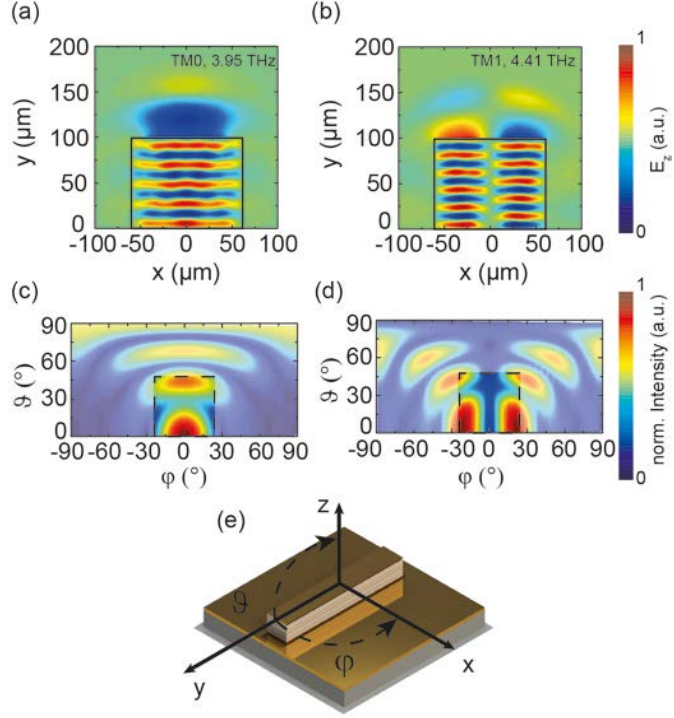

Fig. 2. 3-dimensional finite-element calculations of a Fabry-Pérot cavity. Spatial distribution of the z-component of the electric field $E_z$ in the vertical center (z-direction) of the cavity of a fundamental TM0 mode (a) and the first higher order lateral mode TM1, showing two lateral maxima (b). In (c) and (d), the corresponding far-fields are illustrated. The dashed rectangles indicate the observable section of the far-field with the used measurement setup, which is limited by the parabolic mirror (focal length of 5 cm) which couples the radiation into the spectrometer. (e) Illustration of a double-metal waveguide with etched facets.

For the experimental demonstration, a THz QCL active region, based on the InGaAs/InAlAs material system was used, with a central emission frequency of about 3.8 THz, which was processed into a double-metal waveguide geometry. The devices were realized with etched facets, with a cavity length of 1000 µm, a width of 120 µm and a height of 10 µm. In Fig. 3(a) the emission spectrum, measured with the single pixel DTGS detector is shown. The modes labelled A, B, C and D, are all aligned to a grid with a spacing of 41 GHz, corresponding to a Fabry-Pérot spacing considering a group index of *n=4.4* and a cavity length of 1000 µm. Modes E and F are located in between this grid (however, with the same relative positions within the grid), which consequently correspond to a different type of mode.

The spectrally resolved far-fields are obtained by replacing the single pixel detector with the microbolometer camera. For each individual detector pixel the Fourier transform of the recorded interferogram is calculated. The resulting far-fields are then composed of the spectral components of each detector pixel, corresponding to the frequencies obtained by the integral spectrum in Fig. 3(a). Measurements of the modes A, B, C and D, Fig. 3(b), show one prominent maximum in the center of the far-field, while E and F exhibit a minimum at this position. This is in accordance with the simulations presented in Fig. 2(c) and (d), and we thus attribute modes A, B, C and D to be of TM0, and E and F to be of TM1 type. It may be mentioned that the tilted far-field compared to the simulation is due to possible misalignments of the optical arrangement inside the FTIR spectrometer, which cannot be accessed.

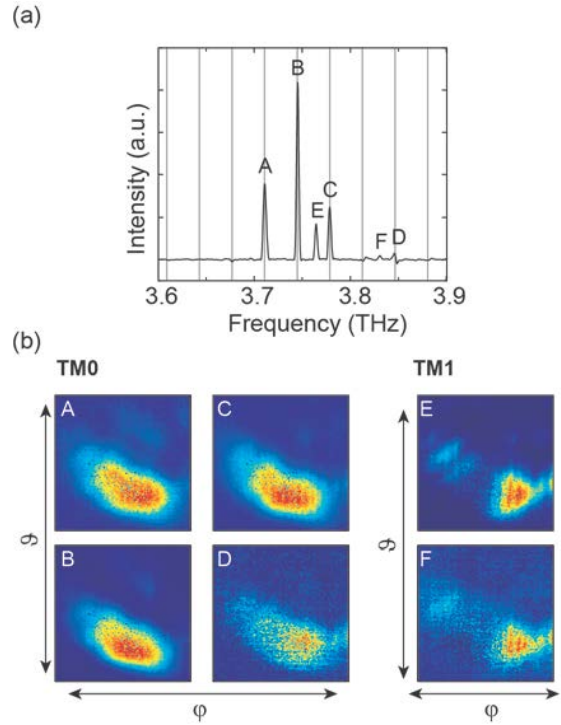

Fig. 3. Spectrally resolved far-field measurements of a laser with a Fabry-Pérot cavity. (a) Spectrum of the laser obtained using a single pixel DTGS detector. The vertical lines indicate the Fabry-Pérot spacing for a cavity with a length of 1000 µm and a group index of n = 4.4 (representing the InGaAs active region). Modes labelled A, B, C and D perfectly align with this grid and thus correspond to the same mode type. E and F are located in between, though at the same relative positions with respect to the grid, and thus belong to a different mode type. (b) A single lobed far-field is observed for modes A, B, C and D, which thus are attributed to be of TM0 type. Far-fields corresponding to E and F show 2 lobes with a minimum at the center, as expected for TM1 modes. The measurements are thus in good agreement with the simulations presented in Fig. 2.

### 3.2 Far-Field of a Random Laser Cavity

Random laser cavities provide an unconventional method to achieve broadband lasing emission at THz frequencies whereby some essential drawbacks of competing techniques can be circumvented. The devices investigated here consist of randomly placed holes, acting as scattering elements, which provide the necessary feedback for laser operation [14]. The particular scattering configuration determines multiple laser modes with different spatial mode profile within the cavity. This multi-frequency feedback mechanism thus enables broadband laser emission [20]. Moreover, the individual holes couple out the radiation in vertical direction (perpendicular to the laser surface). Previous studies revealed that most modes lead to a near-field net-polarization on top of the cavity, leading to constructive interference of the vertically emitted radiation [14]. For this reason, a large fraction of the emission is directed in vertical direction. The individual random laser modes exhibit different far-field profiles adding up for multimode emission. Furthermore, with an adequately chosen scattering density (filling fraction), an output beam with nearly diffraction limited emission profile is obtained [14]. Thus, random laser cavities can be used to achieve both broadband, as well as highly directional emission, which is not possible with competing cavity concepts. To better understand the underlying mechanism and to further optimize these resonators, a spectrally resolved far-field study is necessary, making the presented interferometric technique ideally suited for this objective.

The actual devices consist of a disk-shaped resonator (500 µm diameter) with double-metal waveguide (10 µm height). Randomly placed holes (20 µm diameter) are defined by dry etching through the whole active region, see Fig. 4(a). The spectrum measured in surface direction of a particular random laser with 34% filling fraction (34% of the laser active region is perforated by holes) is depicted in Fig. 4(b), showing unequally spaced modes within the whole gain bandwidth of the laser (~300 GHz). Spectrally resolved far-fields corresponding to the laser modes in Fig. 4(b) are depicted in Fig. 4(c). Each mode exhibits an individual emission profile. However, it can be seen that several far-fields show components radiating in the same direction (see e.g. the far-field of mode A, D, E, F), which add up in the total spectrally integrated far-field. To verify the performed measurements an incoherent superposition of all extracted far-fields, depicted in Fig. 4(c), is compared to a spectrally integrated far-field measurement of the same device, which is obtained by a scanning single pixel DTGS detector, mounted on a translation stage 6 cm in front of the laser, see Fig. 4(d) and (e), respectively. The measurements show a very good accordance, thus proving that the presented method is indeed suitable to investigate the spectrally resolved far-field characteristics of multimode laser cavities.

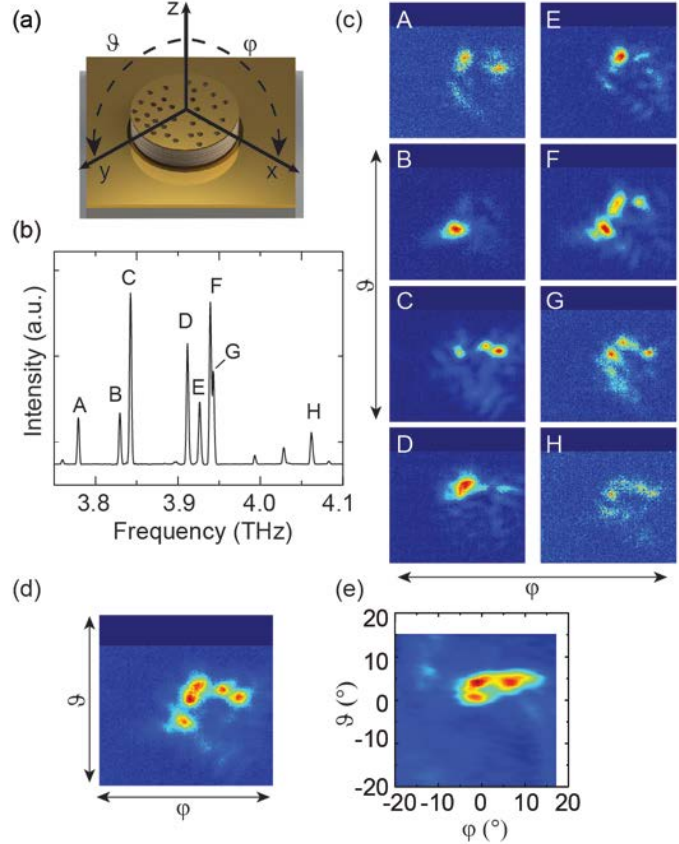

**Fig. 4.** (a) Illustration of the investigated random laser cavity, consisting of a circular resonator with randomly placed holes, which act as scattering- and outcoupling elements. (b) Measured spectrum of a device with 34% filling fraction, obtained using a single pixel DTGS detector. (c) Spectrally resolved far-fields corresponding to the modes in (b). Although each mode shows an individual emission pattern, some components radiate in the same direction. (d) Incoherent superposition of the far-fields in (c), showing a very good agreement with the spectrally integrated measurements depicted in (e), obtained using a single pixel detector, which is scanned at a distance of 6 cm in front of the device.

## 4. Conclusion

In conclusion we performed spectrally resolved far-field measurements on multimode THz QCLs with different resonator types using a microbolometer focal plane array in combination with an FTIR spectrometer. The measurement is performed in free-running mode, providing very short acquisition times. Moreover, the spectral resolution is the same as for measurements obtained using the internal single-pixel detector since it depends on the scanning path of the moving interferometer mirror only. With the presented method we can clearly distinguish between simultaneously occurring fundamental TM0 and higher order lateral TM1 modes of Fabry-Pérot type cavities. Furthermore, we analyzed the far-fields of the modes of a random laser cavity, which show completely different emission characteristics. We find that some modes partially radiate in the same direction, which can lead to a very directional far-field for an adequately chosen random pattern [14]. Thus, using this technique the emission mechanism of such cavities can be further investigated, which is very important for the future optimization of the far-field and for the development of broadband cavities in general.

Moreover, we demonstrate that combining a microbolometer camera with an FTIR spectrometer can be extended from the mid-infrared to the THz regime. In combination with a broadband QCL light source, this could pave the way for future applications relying on two dimensional spatially resolved spectroscopy in this spectral region.

The authors acknowledge financial support received by the Austrian Science Fund (FWF) through Projects No. F49 (SFB NextLite), No. W1210 (DK CoQuS), and No. W1243 (DK Solids4Fun). H.D. is an APART Fellow of the Austrian Academy of Sciences.